\documentclass[11pt,a4paper]{article}
\usepackage[utf8]{inputenc}
\usepackage{a4wide}
\usepackage{amsmath}
\usepackage{cite}
\usepackage{xcolor}
\usepackage{amssymb}
\usepackage{amsfonts}
\usepackage{booktabs}
\usepackage{makecell}
\usepackage{multirow}
\usepackage{pdflscape}
\usepackage{textcomp}
\usepackage{gensymb}
\usepackage{bigstrut}
\usepackage{array}
\usepackage{enumerate}
\usepackage[small,bf]{caption}
\usepackage{graphicx}
\usepackage{hyperref}
\usepackage{mathbbol}
\usepackage{amssymb}
\usepackage{appendix}
\usepackage{verbatim}
\usepackage{geometry}
\usepackage{makecell}
\usepackage{listings}
\usepackage{mathtools}
\usepackage{dsfont}
\usepackage{arydshln}
\usepackage{subcaption}
\usepackage{float}
\usepackage{afterpage}
\usepackage{tikz}

\usetikzlibrary{fit,matrix}

\usepackage{amsmath}

\DeclareMathOperator{\diag}{diag}
\DeclareMathOperator{\im}{Im}

\newcommand{\id}{\mathds{1}}

\allowdisplaybreaks
\numberwithin{equation}{section}

\begin{document}

\begin{titlepage}

\vspace*{-15mm}
\begin{flushright}
\end{flushright}
\vspace*{5mm}

\begin{center}
{\bf\LARGE 
Addressing the CKM Unitarity Problem\\[3mm] with a Vector-like Up Quark} \\[8mm]
G.~C.~Branco  \footnote{E-mail: \texttt{gbranco@tecnico.ulisboa.pt}},
J.~T.~Penedo  \footnote{E-mail: \texttt{joao.t.n.penedo@tecnico.ulisboa.pt}},
Pedro~M.~F.~Pereira  \footnote{E-mail: \texttt{pedromanuelpereira@tecnico.ulisboa.pt}},\\[1mm]
M.~N.~Rebelo  \footnote{E-mail: \texttt{rebelo@tecnico.ulisboa.pt}},
and
J.~I.~Silva-Marcos \footnote{E-mail: \texttt{juca@cftp.tecnico.ulisboa.pt}}\\
\vspace{5mm}
Centro de
F{\'\i}sica Te\'orica de Part{\'\i}culas, CFTP,  Departamento de
F\'{\i}sica,\\ {\it Instituto Superior T\'ecnico, Universidade de Lisboa, }
\\ {\it Avenida Rovisco Pais nr. 1, 1049-001 Lisboa, Portugal;}
\end{center}
\vspace{2mm}

\begin{abstract}
We point out that hints of deviations from unitarity in the first row of the CKM matrix may be explained by the presence of a single vector-like top. We study how the stringent experimental constraints arising from CP Violation in the kaon sector and from meson mixing such as $D^0$-$\overline{D}^0$, $K^0$-$\overline{K}^0$ and $B^0_{d,s}$-$\overline{B}^0_{d,s}$ can be satisfied in the proposed framework. 
In order for the deviations from unitarity to be of the required size while keeping the theory perturbative, the new top quark should have a mass $m_T \lesssim 7$ TeV which could be probed in upcoming experiments at the energy frontier.
\end{abstract}

\end{titlepage}
\setcounter{footnote}{0}
%

\section{Introduction}
\label{sec:intro}

Unitarity of the Cabibbo-Kobayashi-Maskawa (CKM) matrix is an important feature of the 
Standard Model (SM) which has to be tested experimentally. Given the present experimental precision, one of the predictions which can be tested to a high degree of accuracy 
is the normalisation of the first row of the CKM matrix. This is the row that has been measured with better precision.

Recent measurements of $|V_{us}|$ and $|V_{ud}|$ indicate that unitarity of the first row may be violated,
$|V_{ud}|^2+|V_{us}|^2+|V_{ub}|^2 < 1$, at the level of two or three standard deviations. 
This deficit results from new theory calculations of the SM radiative corrections to
$\beta$-decay processes~\cite{Seng:2018yzq,Seng:2018qru}
and was independently confirmed in Refs.~\cite{Czarnecki:2019mwq,Seng:2020wjq,Hayen:2020cxh,Shiells:2020fqp}.
If this result holds, it would be a clear indication for New Physics~\cite{Czarnecki:2004cw,Belfatto:2019swo}.%
\footnote{
This hint stems from tensions between different determinations of the Cabibbo angle (see also~\cite{Grossman:2019bzp}).
Alternative explanations to the Cabibbo angle anomaly not relying on the violation of first-row CKM unitarity are possible~\cite{Bryman:2019bjg}
and have been proposed based on lepton flavour universality violating new physics~\cite{Coutinho:2019aiy,Crivellin:2020lzu}, like extra leptons~\cite{Endo:2020tkb,Crivellin:2020ebi}, extra gauge bosons~\cite{Capdevila:2020rrl}, or a charged scalar singlet~\cite{Crivellin:2020oup,Crivellin:2020klg,Felkl:2021qdn}, and from the perspective of effective field theory~\cite{Kirk:2020wdk,Alok:2020jod,Crivellin:2021njn,Crivellin:2021rbf}.}
The fact that a unitary CKM matrix has been so successful up till now in accommodating a vast number of experimental data both on quark mixing and CP violation indicates that deviations from $3 \times 3$ unitarity, if present, should be small. Extensions of the SM with the addition of vector-like quarks are minimal and have the notable feature of leading to naturally suppressed violations of  $3 \times 3$  unitarity as well as to naturally suppressed flavour changing neutral currents (FCNC) at tree level. Early references include~\cite{delAguila:1982fs,Branco:1986my,Langacker:1988ur,Barbieri:1988im,Nir:1990yq,Bento:1991ez,Nardi:1991rg,Branco:1992wr,Branco:1992uy,delAguila:1997vn,Barenboim:1997pf,Barenboim:1997qx,delAguila:2000aa,delAguila:2000rc}.

Recently it has been suggested~\cite{Belfatto:2019swo} that the addition of a down-type ($Q =-1/3$) vector-like isosinglet quark may lead to deviations from unitarity capable of accommodating the recent measurements of $|V_{us}|$ and $|V_{ud}|$.
In this paper, we point out that such deviations from unitarity in the first row of the CKM can alternatively be explained through the introduction of an up-type ($Q=2/3$) isosinglet quark.
This solution is especially interesting because the experimental limits on FCNC in the up sector are less stringent than those in the down sector. 
Furthermore, as explained in section~\ref{sec:phenomenology}, we find that this solution may be more plausible than the addition of a down-type vector-like quark.

The key question addressed in this paper is whether it is possible 
to have the required deviations from unitarity while at the same time conforming to 
the stringent experimental constraints arising from $D^0$-$\overline{D}^0$,
$K^0$-$\overline{K}^0$ and  $B^0_{d,s}$-$\overline{B}^0_{d,s}$ mixings and to the requirement of having perturbative Yukawa couplings. We will show that this is indeed the case and furthermore a scenario with an extra vector-like up quark also provides exciting prospects for New Physics at the reach of the LHC and its next upgrade.

The paper is organised as follows. In the next section we present our notation and describe the framework of our extension of the SM, including the pattern of non-unitary mixing and $Z$
and Higgs FCNC. We use the fact that one can always choose, without loss of generality, a weak basis (WB) where the down mass matrix is already real and diagonal and as a result the physical mixing matrix can be read off  directly from the left-hand side unitary matrix that diagonalises the up mass matrix. This choice of WB significantly simplifies the search for allowed regions of parameter space, due to the reduction of free parameters it implies.

In section~\ref{sec:newphysics} we present the prospects for New Physics, including new contributions to meson mixing, indirect CP violation in $K_L \rightarrow \pi \pi$ as well as rare top decays $t \to q Z$ ($q=u,c$). In section~\ref{sec:phenomenology} we present the results of our numerical analysis and we give an explicit benchmark. Finally in the last section we present our conclusions.

\section{Notation and Framework}
\label{sec:setup}

\subsection{Lagrangian and mass matrices}
\label{sec:lagrangian}
%
We consider the SM with the minimal addition of one up-type ($Q=+2/3$) isosinglet vector-like quark (VLQ), denoted $T_{L}^0$ and $T_{R}^0$, which transforms as a triplet under $SU(3)_c$. 
The SM scalar sector remains unchanged. The relevant part of the Lagrangian reads, in the flavour basis:
\begin{equation}
\begin{aligned}
-\mathcal{L}_u \,\,\supset\,\,\,
  &Y^u_{ij} \,\,\overline{Q}^0_{Li}\,\tilde\phi\,u_{Rj}^0
\,+\, \overline{Y}_{i} \,\,\overline{Q}^0_{Li}\,\tilde\phi\,T_{R}^0\\[1mm]
\,+\,\, &\overline{M}_{i} \,\,\overline{T}_{L}^0\,u_{Ri}^0
\,+\, M \,\,\overline{T}_{L}^0\,T_{R}^0
\,+\, \text{h.c.}
\,,
\label{eq:lagrangean}
\end{aligned}
\end{equation}
where $Y^u$ are the SM Yukawa couplings, $\phi$ denotes the Higgs doublet ($\tilde\phi =\epsilon \,\phi^*$), $Q^0_{Li} = \left(u_{Li}^0\,\,d_{L i}^0\right)^T$ and $u_{R i}^0$ ($i,j = 1,2,3$) denote the SM quark doublets and up-type quark singlets, respectively.
Here, $\overline{Y}$ denotes Yukawa couplings to the extra right-handed field, while $\overline{M}$ and $M$ correspond, at this level, to bare mass terms.
The down-sector Yukawa Lagrangian is simply $-\mathcal{L}_d \,=\, Y^d_{ij} \,\overline{Q}^0_{Li}\,\phi\,d_{Rj}^0+ \text{h.c.}\,$.
Note that the right-handed VLQ field $T_R^0$ is a priori indistinguishable from the SM fermion singlets $u_{Ri}^0$, since they possess the same quantum numbers.

Following the spontaneous breakdown of electroweak symmetry,
the terms in the first line of eq.~\eqref{eq:lagrangean} give rise 
to the $3\times 3$ mass matrix $m = \frac{v}{\sqrt{2}}\,Y^u$ and 
to the $3 \times 1$ mass matrix $\overline{m}= \frac{v}{\sqrt{2}}\,\overline{Y}$ for the up-type quarks,
with $v \simeq 246$ GeV.
Together with $\overline{M}$ and $M$, they make up the full $4\times 4$ mass matrix $\mathcal{M}_u$,
\begin{align}
- \mathcal{L} \,\,\supset\,\, 
\begin{pmatrix}
\overline{u}_L^0 & \overline{T}_L^0 
\end{pmatrix}\,
\mathcal{M}_u\,
\begin{pmatrix}
u_R^0 \\[2mm] T_R^0
\end{pmatrix} 
\,+\,
\overline{d}_L^0 \,\mathcal{M}_d\, d_R^0
\,+\, \text{h.c.}\,,
\end{align}
with
\begin{align}
\renewcommand{\arraystretch}{1}
\setlength{\extrarowheight}{6mm}
\mathcal{M}_u \,=\,
\left(\begin{array}{c;{2pt/2pt}c}
\quad\,\, m \hphantom{\quad\,\,\,}& \overline{m}\!\!
\\[5mm] \hdashline[2pt/2pt] \\[-1.5cm]
{\quad\,\,\overline{M} \hphantom{\quad\,\,\,}} &
{M\!}
\end{array}\right) \,.
\label{eq:genmass}
\end{align}
It is important to emphasize that, in general, $\mathcal{M}_{u}$ is not symmetric nor Hermitian and that a hierarchy $\overline{M} \sim M \gg \overline{m} \sim m$ is expected.
One is allowed, without loss of generality, to work in a weak basis (WB) where the $3 \times 3$ down-quark mass matrix $\mathcal{M}_d = \frac{v}{\sqrt{2}}\, Y^d$ is diagonal. In what follows we take $\mathcal{M}_d = \mathcal{D}_d =\diag(m_d,m_s,m_b)$.

The matrix
$\mathcal{M}_u$ can be diagonalised by bi-unitary transformations (their singular value decompositions) as
\begin{align}
\renewcommand{\arraystretch}{1}
\setlength{\extrarowheight}{6pt}
\allowdisplaybreaks[0]
    \mathcal{V}_L^\dagger
    \,\mathcal{M}_u\,
    \mathcal{V}_R
    \,=\, \mathcal{D}_u\,,
\label{eq:diagonalisation}
\end{align}
with $\mathcal{D}_u = \diag(m_u,m_c,m_t,m_T)$, where $m_T$ is the mass of the new and heavy up-type quark $T$. The unitary rotations $\mathcal{V}_{L,R}$ relate the flavour basis to the physical basis.

\subsection{Parameterisation}
\label{sec:param}
%
It is convenient to define $3 \times 4$ matrices $A_{L,R}$ as the first three rows of $\mathcal{V}_{L,R}$, denoting the remaining fourth row by $B_{L,R}$,
\begin{align}
\renewcommand{\arraystretch}{1.2}
\mathcal{V}_{L,R}
\,\equiv\,
\setlength{\extrarowheight}{1.2pt}
    \left(\begin{array}{c}
     { }\\[-2mm]
      \qquad A_{L,R} \qquad 
      \\[4mm] \hdashline[2pt/2pt]
      \qquad B_{L,R} \qquad  { }
    \end{array}\right) 
\,.
\label{eq:A,B}
\end{align}
For a fixed index ($L$ or $R$, omitted) the unitarity of $\mathcal{V}$ implies
$AA^\dagger = \id_{3 \times 3}$, $BB^\dagger = 1$, $AB^\dagger = 0$, $BA^\dagger = 0$ and $A^\dagger A+ B^\dagger B = \id_{4 \times 4}$.

In this context, it is convenient to parameterise $\mathcal{V}_L^\dagger$ instead of $\mathcal{V}_L$, since the physical mixing matrix can be read off directly from the former, as we will see shortly. We consider the following parameterisation in terms of 6 mixing angles and 3 phases~\cite{Botella:1985gb}:
\begin{equation}
\begin{aligned}
    \mathcal{V}_L^\dagger = &
\begin{pmatrix}
 1 & 0 & 0 & 0 \\
 0 & 1 & 0 & 0 \\
 0 & 0 & c_{34} & s_{34} \\
 0 & 0 & -s_{34} & c_{34}
\end{pmatrix}
\begin{pmatrix}
 1 & 0 & 0 & 0 \\
 0 & c_{24} & 0 & s_{24}  e^{-i \delta_{24}}  \\
 0 & 0 & 1 & 0 \\
 0 & - s_{24} e^{i \delta_{24}} & 0 & c_{24} 
\end{pmatrix}
\begin{pmatrix}
 c_{14} & 0 & 0 & s_{14} e^{-i \delta_{14}} \\
 0 & 1 & 0 & 0 \\
 0 & 0 & 1 & 0 \\
 - s_{14}  e^{i \delta_{14}}& 0 & 0 & c_{14} 
\end{pmatrix}
\\
&\begin{pmatrix}
 1 & 0 & 0 & 0 \\
 0 & c_{23} & s_{23} & 0 \\
 0 & -s_{23} & c_{23} & 0 \\
 0 & 0 & 0 & 1
\end{pmatrix}
\begin{pmatrix}
 c_{13} & 0 & s_{13} e^{-i\delta_{13}} & 0 \\
 0 & 1 & 0 & 0 \\
 - s_{13} e^{i \delta_{13}}  & 0 & c_{13} & 0 \\
 0 & 0 & 0 & 1
\end{pmatrix}
\begin{pmatrix}
 c_{12} & s_{12} & 0 & 0 \\
 -s_{12} & c_{12} & 0 & 0 \\
 0 & 0 & 1 & 0 \\
 0 & 0 & 0 & 1
\end{pmatrix}
\,,
\end{aligned}
\label{eq:Vparam}
\end{equation}
where $c_{ij} = \cos \theta_{ij}$ and $s_{ij} = \sin \theta_{ij}$, with $\theta_{ij} \in [0,\pi/2]$, $\delta_{ij} \in [0,2\pi]$.

\subsection{Non-unitary mixing}
\label{sec:NUmixing}
%
The interactions of SM quarks with the $W$, $Z$ and Higgs bosons are modified in the presence of VLQs. Going from the flavour basis to the physical basis, the charged current Lagrangian becomes
\begin{equation}
\begin{aligned}
    \mathcal{L}_W \,=\, &- \frac{g}{\sqrt{2}}
    \,\overline{u}_{Li}^0\,
    \gamma^\mu\,
    d_{Li}^0\,
    W_\mu^+
    \,+\,\text{h.c.}
    \\
    \,\rightarrow\,\,\,
     &- \frac{g}{\sqrt{2}}
    \begin{pmatrix}
\overline{u}_L & \overline{T}_L 
\end{pmatrix}\,
V\,    \gamma^\mu
\,d_L\,
    W_\mu^+
    \,+\,\text{h.c.}
    \,,
    \label{eq:LCC}
\end{aligned}
\end{equation}
where one identifies an enlarged $4\times 3$ mixing matrix $V$, corresponding to the first three columns of $\mathcal{V}_L^\dagger$, namely
\begin{align}
    V = A_L^\dagger = \left(\begin{array}{ccc}
c_{12} \,c_{13}\, c_{14} & s_{12} \,c_{13} \,c_{14} &  s_{13} \,c_{14}\, e^{-i \delta_{13}}\\
 \ldots & \ldots & \ldots \\
 \ldots & \ldots & \ldots \\[1.5mm]
       \hdashline[2pt/2pt]      
 \ldots & \ldots & \ldots \\[1.5mm]
    \end{array}\right) \equiv  \left(\begin{array}{c}
     { }\\[-2mm]
      \qquad K_\text{CKM} \qquad 
      \\[4mm] \hdashline[2pt/2pt]
      \qquad K_T \qquad  { }
    \end{array}\right) 
    \,.
    \label{eq:V}
\end{align}
The first three rows of $V$, collectively denoted $K_\text{CKM}$, play the role of the $3\times 3$ Cabibbo-Kobayashi-Maskawa (CKM) mixing matrix.

It is clear from our choice of parameterisation that in the limit of $\theta_{14}, \theta_{24}, \theta_{34}$ going to zero there is no mixing with the new quark, $K_\text{CKM}$ is unitary and its parameterisation reduces to the standard one~\cite{Zyla:2020zbs}, while $K_T = (0,0,0)$.
However, it is crucial to note that in general $K_\text{CKM}$ is not unitary.
While $V^\dagger V = \id$, one generically has $V V^\dagger \neq \id$ and a violation of first-row CKM unitarity, $|V_{ud}|^2+|V_{us}|^2+|V_{ub}|^2 < 1$, is possible.
In the parameterisation we are using, the deviations from unitarity $\Delta_n$ of the $n$-th row of the quark mixing matrix $V$ take a simple form:
\begin{equation}
\begin{aligned}
\Delta \,\equiv\, \Delta_1 \,&=\, 1 - |V_{ud}|^2  - |V_{us}|^2 - |V_{ub}|^2 \,&\!\!\!\!=\, \left|\mathcal{V}_{L_{41}}^*\right|^2 \,&=\, s_{14}^2 \,,\\[1mm]
\Delta_2 \,&=\, 1 - |V_{cd}|^2  - |V_{cs}|^2 - |V_{cb}|^2 \,&\!\!\!\!=\, \left|\mathcal{V}_{L_{42}}^*\right|^2 \,&=\, c_{14}^2 \,s_{24}^2 \,,\\[1mm]
\Delta_3 \,&=\, 1 - |V_{td}|^2  - |V_{ts}|^2 - |V_{tb}|^2 \,&\!\!\!\!=\, \left|\mathcal{V}_{L_{43}}^*\right|^2 \,&=\, c_{14}^2\, c_{24}^2\, s_{34}^2 \,.
\label{eq:defdelta}
\end{aligned}
\end{equation}
The experimental data suggests $\sqrt{\Delta} \sim 0.04$~\cite{Belfatto:2019swo}.
In order to isolate the
deviations from unitarity, one can also
consider the left-polar decomposition
\begin{equation}
    K_\text{CKM}
    \,=\, H_L\, U_\text{CKM} \,\equiv\, (\id - \eta)\, U_\text{CKM}\,,
    \label{eq:polar}
\end{equation}
where $U_\text{CKM}$ is a unitary matrix and $H_L$ and $\eta$ are Hermitian matrices. The matrix $\eta$ can be written in terms of the $\Delta_i$, see eq.~\eqref{eq:eta}. 
It is known that in the present framework deviations from unitarity are naturally suppressed by the ratios $m_q/m_T$ ($q=u,c,t$)~\cite{delAguila:1982fs,Branco:1986my,delAguila:1987nn,Langacker:1988ur,Cheng:1991rr}, as we discuss in section~\ref{sec:perturbativity}.

\subsection{\texorpdfstring{$Z$}{Z}- and Higgs-mediated FCNCs}
\label{sec:ZHFCNC}
%
Changing to the physical basis, the neutral current Lagrangian becomes:
\begin{equation}
\begin{aligned}
    \mathcal{L}_Z \,=\, &- \frac{g}{c_W}
    \,
    \bigg[
       \frac{1}{2} \left(\overline{u}_{Li}^0\,\gamma^\mu\,u_{Li}^0
      - \overline{d}_{Li}^0\,\gamma^\mu\,d_{Li}^0 \right)
      \\&\qquad\quad\,
      -\frac{2}{3}s_W^2 \left(
            \overline{u}_{i}^0\,\gamma^\mu\,u_{i}^0
           +\overline{T}^0\,\gamma^\mu\,T^0
                        \right)
      +\frac{1}{3}s_W^2 \left(
            \overline{d}^0_{i}\,\gamma^\mu\,d_{i}^0
                        \right)
    \bigg]\,Z_\mu
    \\[2mm]
    \,\rightarrow\,\,\,
     &-  \frac{g}{c_W}
    \,
    \bigg[
       \frac{1}{2} 
           \begin{pmatrix} \overline{u}_L & \overline{T}_L \end{pmatrix}\,
F^u\, \gamma^\mu \begin{pmatrix} u_L \\[2mm] T_L \end{pmatrix} 
-\frac{1}{2}
            \overline{d}_{Li}\,
 \gamma^\mu d_{Li}
      \\&\qquad\quad\,
      -\frac{2}{3}s_W^2 \left(
            \overline{u}_{i}\,\gamma^\mu\,u_{i}
           +\overline{T}\,\gamma^\mu\,T
                        \right)
      +\frac{1}{3}s_W^2 \left(
            \overline{d}_{i}\,\gamma^\mu\,d_{i}
                        \right)
    \bigg]\,Z_\mu
    \,,
\end{aligned}
\label{eq:LZ}
\end{equation}
where $s_W$ and $c_W$ are respectively the sine and cosine of the Weinberg angle. We have further defined the complete spinors $\psi = \psi_L + \psi_R$, with $\psi \in \{u^{(0)},d^{(0)},T^{(0)}\}$.
The structure of the second line in eq.~\eqref{eq:LZ} is invariant under the rotation to the physical basis. However, the first line is modified, showing that the presence of the VLQ singlet generically brings about a violation of the GIM mechanism~\cite{Glashow:1970gm},
leading to tree-level $Z$-mediated flavour changing neutral currents (FCNC)~\cite{delAguila:1982fs,Branco:1986my}.
In particular, $F^u$ is a $4\times 4$ Hermitian matrix,
\begin{align}
    F^u =\, V V^\dagger\, \,=\, A_L^\dagger A_L
    \,=\, \id - B_L^\dagger B_L\,,
\label{eq:Wu}
\end{align}
where $V$ is the $4\times 3$ matrix appearing in the charged currents.
Using eq.~\eqref{eq:V}, one finds
\begin{align}
\renewcommand{\arraystretch}{1}
\setlength{\extrarowheight}{6mm}
    F^u\,=\,
\left(\begin{array}{c;{2pt/2pt}c}
\quad\,\, K_\text{CKM}K_\text{CKM}^\dagger \hphantom{\quad\,\,\,}& K_\text{CKM} K_T^\dagger\!\!
\\[5mm] \hdashline[2pt/2pt] \\[-1.5cm]
{\quad\,\,K_T K_\text{CKM}^\dagger \hphantom{\quad\,\,\,}} &
{K_T K_T^\dagger\!}
\end{array}\right) \,,
\label{eq:Wu2}
\end{align}
while from eqs.~\eqref{eq:A,B}, \eqref{eq:Vparam} and the last equality of eq.~\eqref{eq:Wu} one explicitly has 
\begin{align}
\label{eq:fu}
\renewcommand{\arraystretch}{1.5}
     F^u  \,&=\,
\begin{pmatrix}
 1 - |\mathcal{V}_{L_{41}}|^2 & -\mathcal{V}_{L_{41}}^* \mathcal{V}_{L_{42}} & -\mathcal{V}_{L_{41}}^* \mathcal{V}_{L_{43}} & -\mathcal{V}_{L_{41}}^* \mathcal{V}_{L_{44}} \\
 -\mathcal{V}_{L_{42}}^* \mathcal{V}_{L_{41}} &  1 - |\mathcal{V}_{L_{42}}|^2  & -\mathcal{V}_{L_{42}}^* \mathcal{V}_{L_{43}} & -\mathcal{V}_{L_{42}}^* \mathcal{V}_{L_{44}} \\
 -\mathcal{V}_{L_{43}}^* \mathcal{V}_{L_{41}} & -\mathcal{V}_{L_{43}}^* \mathcal{V}_{L_{42}} &  1 - |\mathcal{V}_{L_{43}}|^2  &-\mathcal{V}_{L_{43}}^* \mathcal{V}_{L_{44}} \\
 -\mathcal{V}_{L_{44}}^* \mathcal{V}_{L_{41}} & -\mathcal{V}_{L_{44}}^* \mathcal{V}_{L_{42}} & -\mathcal{V}_{L_{44}}^* \mathcal{V}_{L_{43}}&  1 - |\mathcal{V}_{L_{44}}|^2 
\end{pmatrix} \\[1mm] \nonumber
\,&=\,
\small
\begin{pmatrix}
  c_{14}^2
& - c_{14} s_{14} s_{24} e^{i (\delta_{24}-\delta_{14})}
& - c_{14} s_{14} c_{24} s_{34} e^{-i \delta_{14}}
& - c_{14} s_{14} c_{24} c_{34} e^{-i \delta_{14}} \\
  - c_{14} s_{14} s_{24} e^{i (\delta_{14}-\delta_{24})}
& 1 - c_{14}^2 s_{24}^2
& - c_{14}^2 c_{24} s_{24} s_{34} e^{-i \delta_{24}}
& - c_{14}^2 c_{24} s_{24} c_{34} e^{-i \delta_{24}} \\
  - c_{14}   s_{14} c_{24} s_{34} e^{i \delta_{14}}
& - c_{14}^2 c_{24} s_{24} s_{34} e^{i \delta_{24}}
& 1 - c_{14}^2 c_{24}^2 s_{34}^2
& - c_{14}^2 c_{24}^2 c_{34} s_{34} \\
  - c_{14}   s_{14}   c_{24} c_{34} e^{i \delta_{14}} 
& - c_{14}^2 c_{24} s_{24} c_{34}  e^{i \delta_{24}}
& - c_{14}^2 c_{24}^2 c_{34} s_{34}
& 1 - c_{14}^2 c_{24}^2 c_{34}^2
\end{pmatrix}
\\[1mm] \nonumber
\,&\simeq\,
\small
\begin{pmatrix}
 1- \Delta_1
& - \sqrt{\Delta_1} \sqrt{\Delta_2} e^{-i (\delta_{14}-\delta_{24})}
& - \sqrt{\Delta_1} \sqrt{\Delta_3} e^{-i \delta_{14}}
& - \sqrt{\Delta_1} e^{-i \delta_{14}} \\
  - \sqrt{\Delta_1} \sqrt{\Delta_2} e^{i (\delta_{14}-\delta_{24})}
& 1 - \Delta_2
& - \sqrt{\Delta_2} \sqrt{\Delta_3} e^{-i \delta_{24}}
& - \sqrt{\Delta_2} e^{-i \delta_{24}} \\
  - \sqrt{\Delta_1} \sqrt{\Delta_3} e^{i \delta_{14}}
& - \sqrt{\Delta_2} \sqrt{\Delta_3} e^{i \delta_{24}}
& 1 - \Delta_3
& - \sqrt{\Delta_3} \\
  - \sqrt{\Delta_1} e^{i \delta_{14}} 
& - \sqrt{\Delta_2} e^{i \delta_{24}}
& - \sqrt{\Delta_3}
& \Delta_1 + \Delta_2 + \Delta_3
\end{pmatrix}\,.
\end{align}
where in the last equality we have used the definitions~\eqref{eq:defdelta} and the smallness of the $\Delta_i$.

In general, the matrix $F^u$ is not diagonal and thus the model includes potentially dangerous FCNC.
Transitions of the type $u_i \to Z\,u_j$ ($i\neq j$; $i,j=1,\ldots,4$) are controlled by the magnitude of the off-diagonal elements $F^u_{ij}$ (see also the following section).
Using eq.~\eqref{eq:polar} and taking into account the required smallness of the entries of the matrix $\eta$, one further has
\begin{align}
   [ F^u]_{ 3 \times 3} = K_\text{CKM}K_\text{CKM}^\dagger = \id - 2 \eta + \eta^2 \simeq \id - 2 \eta \,,
\label{eq:Wu3}
\end{align}
where we have used eq.~\eqref{eq:polar} and assumed that $\eta$ is small. One then obtains
\begin{align}
\renewcommand{\arraystretch}{1.5}
      \eta \,&\simeq\, \frac{1}{2}
\small
\begin{pmatrix}
  \Delta_1
&  \sqrt{\Delta_1} \sqrt{\Delta_2} e^{-i (\delta_{14}-\delta_{24})}
&  \sqrt{\Delta_1} \sqrt{\Delta_3}e^{-i \delta_{14}}\\
   \sqrt{\Delta_1} \sqrt{\Delta_2} e^{i (\delta_{14}-\delta_{24})}
& \Delta_2
&  \sqrt{\Delta_2} \sqrt{\Delta_3} e^{-i \delta_{24}}\\
   \sqrt{\Delta_1} \sqrt{\Delta_3} e^{i \delta_{14}}
&  \sqrt{\Delta_2} \sqrt{\Delta_3} e^{i \delta_{24}}
& \Delta_3 \\
\end{pmatrix}\,,
\label{eq:eta}
\end{align}
relating the deviations from unitarity of the rows of $K_\text{CKM}$ with the FCNC structure.

\vskip 1mm

The Higgs boson $h$ may likewise mediate tree-level FCNC among up-type quarks, since not all fermions acquire their mass via couplings to the Higgs doublet $\phi$. In the unitary gauge, the interactions of quarks with the Higgs read, first in the flavour basis and then in the physical basis:
\begin{equation}
\begin{aligned}
    \mathcal{L}_h \,=\, &
    -\frac{1}{\sqrt{2}}\,\overline{u}_{Li}^0\big( 
  Y^u_{ij}\,u_{Rj}^0 + \overline{Y}^u_{i}\,T_{R}^0
 \big)h 
 -\frac{1}{\sqrt{2}}\,
  Y^d_{ij}\,\overline{d}_{Li}^0\,d_{Rj}^0\,h 
 \,+\,\text{h.c.}
    \\[2mm]
    \,\rightarrow\,\,\,
     &
     -\begin{pmatrix} \overline{u}_L & \overline{T}_L \end{pmatrix}\,
F^u\, \mathcal{D}_u\begin{pmatrix} u_R \\[2mm] T_R \end{pmatrix} \frac{h}{v}
     - \,\overline{d}_L \, \mathcal{D}_d \, d_R \frac{h}{v}
 \,+\,\text{h.c.}
    \,.
\end{aligned}
\label{eq:Lh}
\end{equation}
Similarly to the case of $Z$-mediated FCNC, the strength of Higgs-mediated FCNC is controlled by the off-diagonal entries of the matrix $F^u$ and by the ratios $m_q/v$, ($q=u,c,t,T$). Note that for transitions involving only the lighter quarks $u$ and $c$, a strong suppression -- by a factor of $m_u/v$ or $m_c/v$ -- is present.

\subsection{Perturbativity}
\label{sec:perturbativity}
%
The magnitude of the first three rows in $\mathcal{M}_u$ is controlled by the scale of electroweak symmetry breaking, $\frac{v}{\sqrt{2}} \simeq 174$ GeV, and capped by the requirement of having perturbative Yukawa couplings $Y^u_{ij}$ and $\overline{Y}_i$ ($i,j=1,2,3$). One can relate the perturbativity condition to quarks masses and deviations $\Delta_i$ from unitarity, obtaining an upper bound on the mass $m_T$ of the new heavy top.
We write the trace of $mm^\dagger +  \overline{m}\, \overline{m}^\dagger$ as
\begin{equation}
  \text{Tr}\,\Big( mm^\dagger +  \overline{m}\, \overline{m}^\dagger \Big)  = p~m_t^2
  \label{eq:treqor}
\end{equation}
where $p$ is a numerical coefficient, constrained by perturbativity.
By taking the trace of $\mathcal{M}_u \mathcal{M}_u^\dagger$, using \eqref{eq:diagonalisation} 
and neglecting $m_{u,c} \ll m_{t,T}$, one finds
\begin{equation}
\begin{aligned}
\text{Tr}\,\Big( mm^\dagger +  \overline{m}\, \overline{m}^\dagger \Big) &\,\simeq\, m_t^2+ m_T^2 -  \Big(\overline{M}\, \overline{M}^\dagger + |M|^2\Big)\\[2mm]
&\,\simeq\, \big(1-|\mathcal{V}_{L_{43}}|^2\big)\,m_t^2 +  \big(1-|\mathcal{V}_{L_{44}}|^2\big)\,m_T^2\\[2mm]
&\,\simeq\, m_t^2 +  \big(\Delta_1+\Delta_2+\Delta_3)\,m_T^2\,,
\end{aligned}
\label{eq:trexp}
\end{equation}
by noting that $1-|\mathcal{V}_{L_{43}}|^2=1-\Delta_3$, with $\Delta_3 \ll 1$, and $1-|\mathcal{V}_{L_{44}}|^2= \sum_i \Delta_i$ (see eq.~\eqref{eq:fu}).
Using eq.~\eqref{eq:treqor} we obtain
\begin{equation} 
\sqrt{\Delta_1 + \Delta_2 + \Delta_3} = \sqrt{p-1}~\frac{m_t}{m_T} ~,
    \label{eq:deltamtmT}
\end{equation}
where it is clear that $p$ should always be bigger than $1$.
Requiring Yukawa couplings of at most $\mathcal{O}(1)$ constrains the last term of eq.~\eqref{eq:trexp} to be of $\mathcal{O}(m_t^2)$. This indicates that  $\sqrt{p-1}$ should also be of $\mathcal{O}(1)$.
Hence, eq.~\eqref{eq:deltamtmT} translates into the approximate upper bound
\begin{equation}
    m_T \lesssim \frac{m_t}{\sqrt{\Delta_1+\Delta_2+\Delta_3}}\,.
    \label{eq:perturbativity}
\end{equation}
Eq.~\eqref{eq:deltamtmT} should come as no surprise as it is well-known that an appealing feature of the present framework is that deviations from unitarity are suppressed by the ratios $m_q/m_T$ ($q=u,c,t$).
In a scenario where $\Delta_{(1)}$ dominates, i.e.~$\Delta \gg \Delta_{2,3}$, one obtains
\begin{equation}
    m_T \lesssim \frac{m_t}{\sqrt{\Delta}}\,,
    \label{eq:perturbativity2}
\end{equation}
which implies $m_T \lesssim 4.4$ TeV for $\sqrt{\Delta}= 0.04$. 
Potentially weaker but more precise bounds may in principle be obtained from perturbative unitarity considerations (see e.g.~\cite{DiLuzio:2016sur}). The derivation of such bounds falls outside the scope of this paper and we make use of the qualitative restriction of eq.~\eqref{eq:perturbativity} in what follows.

\section{New Physics}
\label{sec:newphysics}

A plethora of observables can be modified at the tree or loop level
in the presence of an up-type VLQs (see for instance~\cite{Cacciapaglia:2011fx,Okada:2012gy,Panizzi:2014dwa,Alok:2015iha}). In this work we focus on the New Physics (NP) contributions to neutral meson mixing ($D$, $K$ and $B_{d,s}$ neutral-meson systems), and on the NP enhancement of the rare top decays $t \to q Z$.
The new heavy quark $T$, if light enough, may in principle be produced at present colliders.
The dominant contributions to the pair production cross sections only depend on the mass $m_T$ (see e.g.~\cite{DeSimone:2012fs}). On the other hand, single production mechanisms 
can in part be approximately parameterised via functions of the mixing matrix elements $V_{Td}$, $V_{Ts}$ and $V_{Tb}$ and are more model dependent~\cite{Buchkremer:2013bha}. The same quantities influence the rates of the $T$ decays into vector bosons and SM quarks (see e.g.~\cite{Botella:2016ibj}).
Lower bounds on the VLQ mass, $m_T > 1.3$ TeV~\cite{Aaboud:2018pii} and $m_T \gtrsim 1.0$ TeV~\cite{Sirunyan:2019sza},
have been obtained at the 95\% CL by the ATLAS and CMS collaborations, respectively,
in searches for pair-produced $T$-quarks.%
\footnote{
Mass bounds depend on assumptions on branching ratios (and in the CMS case on the type of analysis: cut-based vs.~neural network).
Experimental searches typically assume that the new quark only couples to the third SM quark generation. 
In the generic case where VLQs mix with all SM quarks, searches may need to be reinterpreted~\cite{Panizzi:2014dwa}.
}

\subsection{\texorpdfstring{$D^0$-$\overline{D}^0$}{D0-D0bar} Mixing}
\label{sec:D0D0bar}
\begin{figure}[t!]
  \centering
    \begin{subfigure}[b]{\textwidth}
  \centering
        \includegraphics[width=0.45\textwidth]{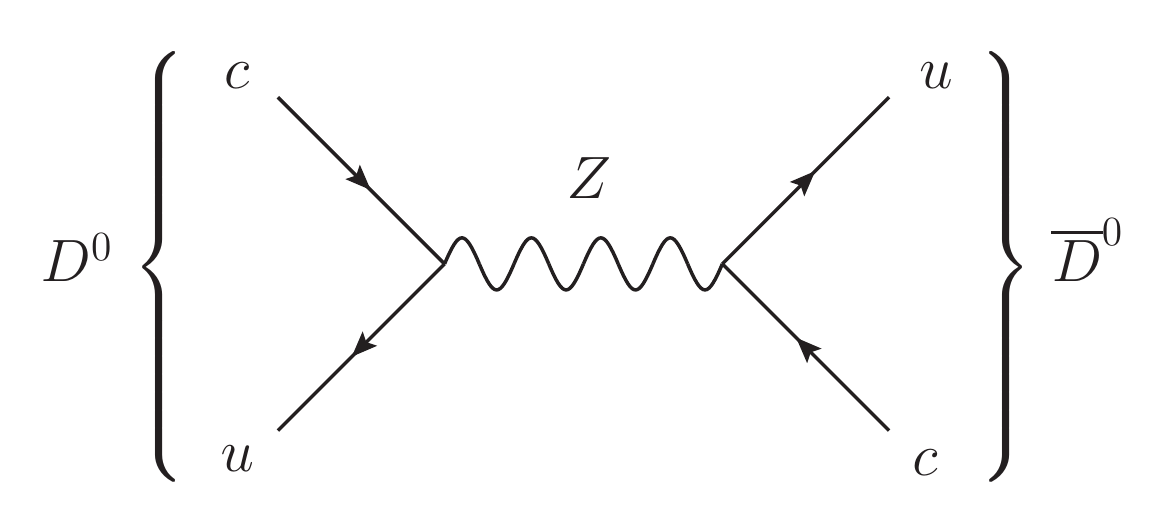}
        \caption{NP contribution to $D^0$-$\overline{D}^0$ mixing via $Z$-mediated FCNC.}
        \label{fig:D}
    \end{subfigure}\\
    \begin{subfigure}[b]{\textwidth}
    \vskip 2mm
  \centering
        \includegraphics[width=\textwidth]{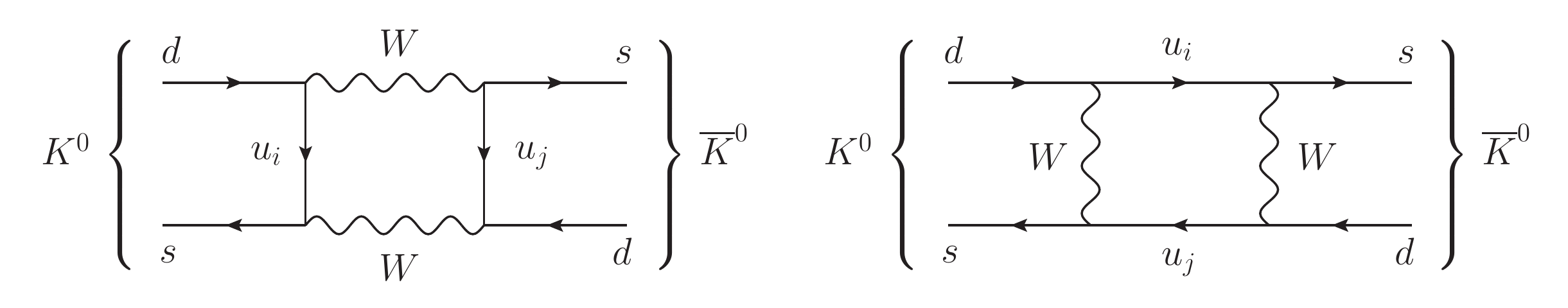}
        \caption{Leading contributions to $K^0$-$\overline{K}^0$ mixing, including the effect of the new heavy quark, $u_{i,j} = u,c,t,T$.}
        \label{fig:K}
    \end{subfigure}\\
    \begin{subfigure}[b]{\textwidth}
    \vskip 2mm
  \centering
        \includegraphics[width=\textwidth]{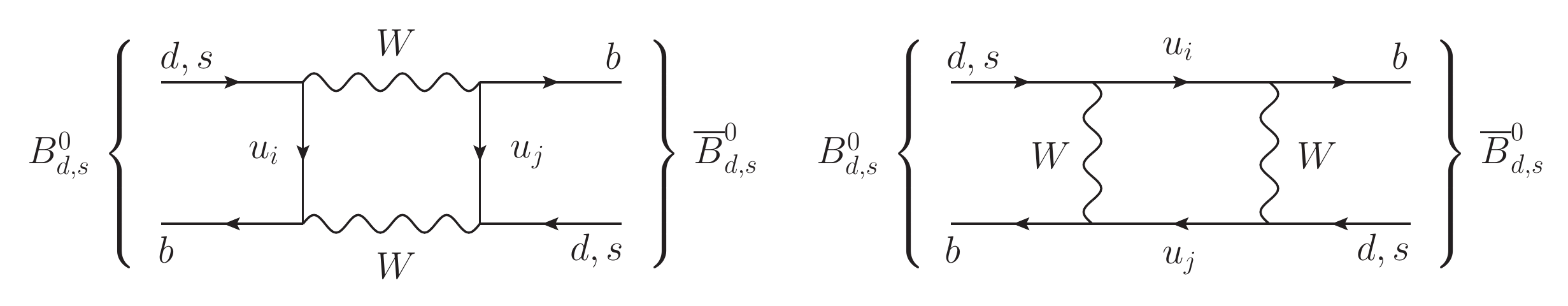}
        \caption{The same as in (b) but for $B_{d,s}^0$-$\overline{B}_{d,s}^0$ mixing.}
        \label{fig:B}
    \end{subfigure}
  \caption{Feynman diagrams for the leading contributions to neutral meson mixing in the presence of one up-type VLQ.}
  \label{fig:diagrams}
\end{figure}

%
The $Z$-mediated tree-level FCNC discussed in the previous section may most notably compete with the SM contribution to $D^0$-$\overline{D}^0$ mixing, which occurs at the loop level~\cite{Branco:1995us}.
The tree-level NP contribution is shown in Figure~\ref{fig:D} and corresponds to the effective Lagrangian
\begin{align}
\mathcal{L}^\text{NP}_\text{eff}
\,=\,
-\frac{G_F}{\sqrt{2}}\left(F^u_{12} \right)^2 (\overline{u}_L \gamma^\mu c_L) (\overline{u}_L \gamma_\mu c_L)
\,,
\end{align}
where $G_F$ is the Fermi constant and $F_{12}^u=-\mathcal{V}_{L_{41}}^* \mathcal{V}_{L_{42}}$.
This $\Delta C = 2$ operator results in a contribution to the $D^0$ mixing parameter $x_D = \Delta m_{D}/\Gamma_D$ of size~\cite{Branco:1995us,Golowich:2009ii}
\begin{align}
x_D^\text{NP}
\,\simeq\, \frac{\sqrt{2}\,m_{D}}{3\, \Gamma_D} \,G_F f_D^2 B_D \,r(m_c,M_Z) \left| \mathcal{V}_{L_{41}}^* \mathcal{V}_{L_{42}} \right|^2\,,
\end{align}
where $m_{D} = 1864.83 \pm 0.05$ MeV, $\Gamma_D = 1/\tau_D$ with
$\tau_D = (410.1 \pm 1.5) \times 10^{-15}$ s~\cite{Zyla:2020zbs}, and the factor $r(m_c,M_Z) \simeq 0.778$ accounts for RG effects, while $B_D \simeq 1.18$~\cite{Buras:2010nd}
and $f_D = 212.0 \pm 0.7$ MeV~\cite{Aoki:2019cca}.

\vskip 2mm
We consider in our analysis the conservative bound $x_D^\text{NP} < x_D^\text{exp}$, where we take $x_D^\text{exp} = 0.39^{+0.11}_{-0.12}\, \%$~\cite{Amhis:2019ckw}. This bound limits from above the product of deviations from unitarity of the first and second rows of the CKM matrix, $\Delta_1\Delta_2 = |\mathcal{V}_{L_{41}}^* \mathcal{V}_{L_{42}}|^2 < 1.5 \times 10^{-8}$. Taking into account the chosen parameterisation, we find $\theta_{14} \theta_{24} \simeq \sqrt{\Delta_1\Delta_2} < 1.2\times 10^{-4}$ in the approximation of small angles $\theta_{14}$ and $\theta_{24}$.

Requiring $x_D^\text{NP} < x_D^\text{exp}$ also keeps under control the NP (tree-level) contribution to the as yet unobserved $\Delta C = 1$ decay $D^0 \to \mu^+\mu^-$~\cite{Golowich:2009ii}. Namely, $\text{Br}(D^0 \to \mu^+\mu^-)_\text{NP} \simeq 3.0 \times 10^{-9}\, x_D^\text{NP} < 1.2 \times 10^{-11}$, while $\text{Br}(D^0 \to \mu^+\mu^-)_\text{exp} < 6.2 \times 10^{-9}$ (90\% CL)~\cite{Zyla:2020zbs}.

\subsection{\texorpdfstring{$K^0$-$\overline{K}^0$}{K0-K0bar} and \texorpdfstring{$B_{d,s}^0$-$\overline{B}_{d,s}^0$}{B0-B0bar} mixing}
\label{sec:downmesons}
%
We now turn to the mixing $N^0$-$\overline{N}^0$ of neutral mesons $N^0 = K^0, B_{(d)}^0, B_s^0$. Given that the valence quarks of these mesons are all of the down type ($K^0 \sim d\bar{s}$, $B^0 \sim d\bar{b}$ and $B_s^0 \sim s\bar{b}$), there is no NP contribution to their mixing at the tree level. Nevertheless, the loop-level short-distance NP contributions may appreciably compete with SM ones. The corresponding diagrams are shown in Figures~\ref{fig:K} and~\ref{fig:B}.
The amplitudes of these box diagrams are proportional to 
$\sum_{i,j} \lambda_i^N \lambda_j^N F(x_i,x_j)$,
where $x_i=(m_i/m_W)^2$ and $F(x_i,x_j)$ is a box loop function~\cite{Branco:1999fs}.
The sum is taken over all up-type quarks ($i,j=u,c,t,T$). One has further defined
\begin{equation}
     \lambda_i^K \,\equiv\, V_{is}^*V_{id}\,,\qquad
     \lambda_i^B \,\equiv\, V_{ib}^*V_{id}\,,\qquad
     \lambda_i^{B_s} \,\equiv\, V_{ib}^*V_{is} \,,
\end{equation}
for each of the considered neutral meson systems.
Unitarity of the columns of $V$, namely $V^\dagger V = \id_{3 \times 3}$, implies 
\begin{equation}
\lambda_u^N+\lambda_c^N+\lambda_t^N+ \lambda_T^N = 0
\end{equation}
and can be used to eliminate the up-quark contributions,
as is typically done in the SM case (see e.g.~\cite{Buras:1997fb}).
The off-diagonal element in the dispersive part of the amplitude for neutral meson mixing is then given by~\cite{Branco:1999fs,Cacciapaglia:2011fx}
\begin{equation}
  \left(M_{12}^N\right)^* \,\simeq\,
  \frac{m_N}{3\sqrt{2}}\, G_F f_N^2 B_N\, \frac{\alpha}{4\pi s_W^2}
  \,\sum_{\substack{i,j=c,t,T}} r_{ij} \,\lambda_i^N\lambda_j^N\, S(x_i,x_j)
  \,,
  \label{eq:M12P}
\end{equation}
where $m_N$, $B_N$ and $f_N$ are the average mass, bag parameter and decay constant of the meson, respectively. Their values are summarised in Table~\ref{tab:mesondata}.%
\footnote{ 
Our conclusions are unchanged when taking into account an updated value for the bag parameter  $B_{B_s}$~\cite{Kirk:2017juj,King:2019lal,DiLuzio:2019jyq}.
}
The factors $r_{ij}$ account for $\mathcal{O}(1)$ QCD corrections to the electroweak diagrams.
Finally, the functions
\begin{align}
S(x_i,x_j)
    &=F(x_i,x_j) -  F(0,x_i) -F(0,x_j) + F(0,0) \nonumber \\
    &= x_i x_j  \bigg[
    \frac{\ln{x_i}}{(x_i-x_j)(1-x_i)^2}\left(1-2x_i+\frac{x_i^2}{4}\right) + (x_i\leftrightarrow x_j) - \frac{3}{4(1-x_i)(1-x_j)}\bigg]\,,
    \nonumber \\[2mm]
    S(x_i) &\equiv \lim_{x_j \rightarrow x_i} S(x_i,x_j) = \frac{x_i}{(1-x_i)^2}\left(1-\frac{11}{4}x_i + \frac{x_i^2}{4}\right)-\frac{3}{2}\frac{x_i^3 \ln{x_i}}{(1-x_i)^3}\,,
\end{align}
are the well-known Inami-Lim functions~\cite{Inami:1980fz}, obeying $S(x_i,x_j) = S(x_j,x_i)$. 
We have taken $x_u \simeq 0$ to a good approximation.
%
\begin{table}
  \centering
  \renewcommand{\arraystretch}{1.5}
  \begin{tabular}{ccccc}
    \toprule
    $N^0$-$\overline{N}^0$ & $m_N$ [MeV] & $\Delta m_N^\text{exp}$ [MeV] & $f_N$ [MeV] & $B_N$  \\
    \midrule
    $K^0$-$\overline{K}^0$     & $497.611\pm 0.013$ & $(3.484 \pm 0.006) \times 10^{-12}$ & $155.7 \pm 0.3$ & $0.717 \pm 0.024$ \\
    $B^0$-$\overline{B}^0$     & $5279.65\pm 0.12 $ & $(3.334 \pm 0.013) \times 10^{-10}$ & $190.0 \pm 1.3$ & $1.30  \pm 0.10 $ \\
    $B_s^0$-$\overline{B}_s^0$ & $5366.88\pm 0.14 $ & $(1.1683 \pm 0.0013) \times 10^{-8}$ & $230.3 \pm 1.3$ & $1.35  \pm 0.06 $ \\
    \bottomrule
  \end{tabular}
  \caption{Mass and mixing parameters~\cite{Zyla:2020zbs} and decay constants and bag parameters~\cite{Aoki:2019cca} for the neutral meson systems with down-type valence quarks considered in section~\ref{sec:downmesons}.}
  \label{tab:mesondata}
\end{table}
%

One can thus isolate the NP contributions to the mass differences $\Delta m_N \simeq 2 |M_{12}^N|$. It follows that
\begin{equation}
\begin{aligned}
     \Delta m_N^\text{NP} &\simeq \frac{G_F^2 M_W^2 m_N f^2_N B_N}{6 \pi^2}
     \left|
       2\, r_{cT} \,\lambda_c^N\lambda_T^N\, S_{cT}
     + 2\, r_{tT} \,\lambda_t^N\lambda_T^N\, S_{tT}
     +   r_{TT} \left(\lambda_T^N\right)^2 S_T \right|\,,
\end{aligned}
\label{eq:deltamP}
\end{equation}
with the shorthands $S_{ij} = S(x_i,x_j)$ and $S_i = S(x_i)$. 

This NP term is sensitive to the mass of the new quark $m_T$ and to the elements of the fourth row of $V$, via $\lambda_T^N$.
In our analysis we take a conservative estimate of the impact of $\Delta m_N^{\text{NP}}$ by requiring $\Delta m_N^\text{NP} < \Delta m_N^\text{exp} $ (we take $r_{iT} = 1$). The experimental values for the $\Delta m_N^\text{exp}$ are given in Table~\ref{tab:mesondata}.

Before proceeding, note that one can obtain some insight into the strength of these constraints by assuming that there are no cancellations and neglecting the charmed term, $S_{cT}\ll S_{tT} < S_T$. Requiring, for illustrative purposes,
\begin{equation}
     \Delta m_N^\text{NP} \sim 
     \frac{G_F^2 M_W^2 m_N f^2_N B_N}{6 \pi^2}
     \left(
     2 \left|\lambda_t^N\right| \left|\lambda_T^N\right|\, S_{tT}
     + \left|\lambda_T^N\right|^2 S_T
     \right)
     < \Delta m_N^\text{exp} 
     \,,
\end{equation}
one obtains approximate $m_T$-dependent bounds on the quantities
\begin{equation}
\begin{aligned}
     \big|\lambda_T^K\big| \,=\, \big|V_{Td}\big| \big|V_{Ts}\big|\,,\quad
     \big|\lambda_T^B\big| \,=\, \big|V_{Td}\big| \big|V_{Tb}\big|\,,\quad
     \big|\lambda_T^{B_s}\big| \,=\, \big|V_{Ts}\big| \big|V_{Tb}\big| \,,
\end{aligned}
\label{eq:boundsP}
\end{equation}
which have no simple expression in the chosen angular parameterisation.
We present such bounds in Table~\ref{tab:mesonbounds} for two benchmark values of the heavy mass, $m_T = 1,3$ TeV.
%
\begin{table}
  \centering
  \renewcommand{\arraystretch}{1.5}
  \begin{tabular}{ccc}
    \toprule
    Observable  & $m_T = 1$ TeV & $m_T = 3$ TeV\\
    \midrule
    $\Delta m_K$ & $\big|V_{Td}\big| \big|V_{Ts}\big| < 7.4 \times 10^{-4}$ & $\big|V_{Td}\big| \big|V_{Ts}\big| < 2.7 \times 10^{-4}$\\
    $\Delta m_B$ & $\big|V_{Td}\big| \big|V_{Tb}\big| < 6.7 \times 10^{-4}$ & $\big|V_{Td}\big| \big|V_{Tb}\big| < 3.4 \times 10^{-4}$\\
    $\Delta m_{B_s}$ & $\big|V_{Ts}\big| \big|V_{Tb}\big| < 3.2 \times 10^{-3}$ & $\big|V_{Ts}\big| \big|V_{Tb}\big| < 1.6 \times 10^{-3}$\\
    $|\epsilon_K|$ & $\big|V_{Td}\big| \big|V_{Ts}\big| \sqrt{|\sin 2 \Theta|} < 8.8 \times 10^{-5}$ & $\big|V_{Td}\big| \big|V_{Ts}\big| \sqrt{|\sin 2 \Theta|} < 3.1 \times 10^{-5}$ \\
    \bottomrule
  \end{tabular}
  \caption{Constraints from neutral meson observables on products of mixing matrix elements ($\Theta = \arg V_{Ts}^* V_{Td}$) for two benchmark masses of the new heavy top quark.}
  \label{tab:mesonbounds}
\end{table}
%

\subsection{CP violation in \texorpdfstring{$K_L \rightarrow \pi \pi$}{KL -> pi pi}}
\label{sec:epsilon}
%

The parameter $\epsilon_K$ describes the indirect CP violation in the kaon system and has been measured to be $|\epsilon_K|^\text{exp} = (2.228 \pm 0.011) \times 10^{-3}$. It can be connected to $M_{12}^K$ via~\cite{Buras:1997fb}
\begin{equation}
|\epsilon_K| = 
\frac{\kappa_\epsilon}{\sqrt{2}\,\Delta m_K} \left|\im M_{12}^K \right|\,,
\end{equation}
where $\kappa_\epsilon \simeq 0.92 \pm 0.02$~\cite{Buras:2008nn}.
Using eq.~\eqref{eq:M12P}, one finds the maximum possible value for the NP contribution,
\begin{equation}
\begin{aligned}
|\epsilon_K|^\text{NP}
    &\,\simeq\, 
\frac{G_F^2 M_W^2\, m_K f^2_K B_K\, \kappa_\epsilon}{12 \sqrt{2}\, \pi^2 \,\Delta m_K}
     \bigg| 
       2\, r_{cT}  S_{cT} \im\big(V_{cs}^*V_{cd}V_{Ts}^*V_{Td}\big) \\
    &\qquad\quad + 2\, r_{tT} S_{tT} \im\big(V_{ts}^*V_{td}V_{Ts}^*V_{Td}\big)
     +   r_{TT} S_T \im\Big[\big(V_{Ts}^{*} V_{Td}\big)^2\Big]  \bigg|\,.
\end{aligned}
\label{eq:epsilonNP}
\end{equation}
While this expression is manifestly not rephasing-invariant, it holds provided $\lambda_u^K$ is real, as is the case for our parameterisation. This maximum NP contribution depends on $m_T$ and on the angles and phases in $V$. In our analysis, we require that its absolute value does not exceed the measured value, $|\epsilon_K|^\text{NP} < |\epsilon_K|^\text{exp}$ (we take $r_{iT} = 1$).

For illustrative purposes, a rough bound on $\lambda^K_T=V_{Ts}^{*} V_{Td}$ can be obtained at the outset by assuming that only the last term in eq.~\eqref{eq:epsilonNP} gives a sizeable contribution, as $S_{cT}\ll S_{tT} < S_T$ and $|V_{td}| |V_{ts}| \sim 3\times 10^{-4}$. Denoting by $\Theta$ the phase of $\lambda^K_T = V_{Ts}^* V_{Td}$, one has
\begin{equation}
|\epsilon_K|^\text{NP}
    \,\sim\, 0.5 \,
\frac{G_F^2 M_W^2\, m_K f^2_K B_K\, \kappa_\epsilon}{12 \sqrt{2}\, \pi^2 \,\Delta m_K}
       S_T \big|V_{Ts}^{*} V_{Td}\big|^2 |\sin 2 \Theta| < |\epsilon_K|^\text{exp}\,,
\label{eq:boundek}
\end{equation}
where the ad-hoc $1/2$ factor takes into account the fact that the $tT$ term may partly cancel the $TT$ one and leads to a more conservative bound. The consequences of eq.~\eqref{eq:boundek} for the previously considered benchmarks ($m_T = 1,3$ TeV) are shown in the last row of Table~\ref{tab:mesonbounds}.

\subsection{Rare top decays \texorpdfstring{$t \to q Z$}{t -> q Z}}
\label{sec:raretop}
%
Provided the new quark mixes with both the light and the third generations, the rates of the rare FCNC decays $t\to q_iZ$ ($q_i=u,c$) may be enhanced with respect to their SM expectation (see e.g.~\cite{Botella:2012ju}). The leading-order NP contribution occurs at tree level and is given by~\cite{AguilarSaavedra:2004wm}
\begin{align}
    \Gamma(t \to q_i Z)_\text{NP} \,\simeq\,
    \frac{\alpha}{32\, s_W^2\, c_W^2}
    \,\left|F^u_{i3} \right|^2\,
    \frac{m_t^3}{M_Z^2} \left(1- \frac{M_Z^2}{m_t^2}\right)^2\left( 1+2\frac{M_Z^2}{m_t^2}\right)
    \,,
\end{align}
with $F_{i3}^u=-\mathcal{V}_{L_{4i}}^* \mathcal{V}_{L_{43}}$ ($i=1,2$). One can approximate the total decay width of the top-quark by $\Gamma_t \simeq \Gamma(t\to b W^+)$
and obtain
\begin{align}
    \text{Br}(t \to q_i Z)_\text{NP} \,\simeq\,
    \frac{\left|\mathcal{V}_{L_{4i}}^* \mathcal{V}_{L_{43}} \right|^2}{2 \left|V_{tb} \right|^2}
    \left(1- \frac{M_Z^2}{m_t^2}\right)^2\!\left( 1+2\frac{M_Z^2}{m_t^2}\right)
    \left(1- 3\frac{M_W^4}{m_t^4} + 2\frac{M_W^6}{m_t^6}\right)^{-1} ,
\end{align}
at leading order. 
This is to be contrasted with the suppressed
Br$(t \to u Z)_\text{SM} \sim 10^{-16}$ and 
Br$(t \to c Z)_\text{SM} \sim 10^{-14}$~\cite{AguilarSaavedra:2004wm} in the SM.
In the small angle approximation, one predicts $\text{Br}(t \to q_i Z)_\text{NP} \simeq 0.46\, \theta_{i4}^2 \,\theta_{34}^2 \sim \Delta_i\, \Delta_3$, which for $\mathcal{O}(0.01)$ angles still exceeds the SM contribution by several orders of magnitude.
At present, the strongest bound on these branching ratios is set by the ATLAS collaboration, namely
Br$(t \to u Z)_\text{exp} < 1.7 \times 10^{-4}$ and 
Br$(t \to c Z)_\text{exp} < 2.4 \times 10^{-4}$ ($95\%$ CL)~\cite{Aaboud:2018nyl}.

As noted in section~\ref{sec:ZHFCNC}, the tree-level NP contribution to the rare decay $t \to q h$ is suppressed with respect to $t\to q Z$ and is not considered in our analysis. The same goes for the new contributions to the rare top decays proceeding at loop level $t \to q g$ and $t \to q \gamma$, which generically exceed the GIM-suppressed SM contributions.

\section{Numerical analysis}
\label{sec:phenomenology}
%

In order to explore the viability of the SM extension with one up-type VLQ we are considering, we have performed a numerical scan of the parameter space of the model. From the outset, the model is constrained by the absolute values of the entries of the CKM matrix. Their present best-fit values, without imposing unitarity, are~\cite{Zyla:2020zbs}
\begin{equation}
    |K_\text{CKM}| = 
    \begin{pmatrix}
      0.97370 \pm 0.00014  & 0.2245 \pm 0.0008 & (3.82 \pm 0.24) \times 10^{-3} \\
      0.221 \pm 0.004      & 0.987 \pm 0.011   & (41.0 \pm 1.4) \times 10^{-3} \\
      (8.0 \pm 0.3) \times 10^{-3} & (38.8 \pm 1.1) \times 10^{-3} & 1.013 \pm 0.030
    \end{pmatrix}\,.
\end{equation}
We further assume~\cite{Botella:2008qm,Botella:2012ju} that the presence of an up-type VLQ does not affect the value of the phase $\gamma = \arg (-V_{ud}V_{cb}V_{ub}^*V_{cd}^*)$, which is obtained from SM tree-level dominated $B$ decays, $\gamma = (72.1^{+4.1}_{-4.5})\degree$~\cite{Zyla:2020zbs}.
We use $N\sigma = \sqrt{\chi^2}$ as a measure of the goodness of fit, where $\chi^2$ is approximated as a sum of priors,
\begin{equation}
    \chi^2 = \sum_{ij} \left(\frac{V_{ij}-V_{ij}^c}{\sigma(V_{ij})}\right)^2 + \left(\frac{\gamma-\gamma^c}{\sigma(\gamma)}\right)^2\,,
\end{equation}
with the superscript $c$ denoting central values and $\sigma(\gamma) = 4.5\degree$. We take $m_c(M_Z) = 0.619 \pm 0.084$ GeV, $m_t(M_Z) = 171.7 \pm 3.0$ GeV~\cite{PhysRevD.77.113016} and require $m_T > 1$ TeV, in line with collider bounds.

The values of the new angles and phases compatible with the above criteria are shown as the dashed regions in the correlation plot of Figure~\ref{fig:correlations}. These regions contract to the solid green $2\sigma$ and $3\sigma$ contours after all the constraints from the previous sections are taken into account. These constraints comprise the bounds on $x_D$, $\Delta m_N$ ($N=K^0, B^0, B^0_s$), and $|\epsilon_K|$ discussed in section~\ref{sec:newphysics} and the perturbativity bound of eq.~\eqref{eq:perturbativity}.
One sees that relatively large values for both $\theta_{14} \simeq \sqrt{\Delta}$ and $\theta_{34}$ are preferred by the data, which disfavour $\theta_{34}=0$ at more than $2\sigma$. Conversely, $\theta_{24}$ is compatible with zero and the preference shown for small values of this angle is driven by the constraint coming from $D^0$-$\overline{D}^0$ mixing (see section~\ref{sec:D0D0bar}).

\afterpage{%
\begin{figure}[h!]
  \centering
\makebox[0pt]{\begin{minipage}{1.15\textwidth}
  \includegraphics[width=\textwidth]{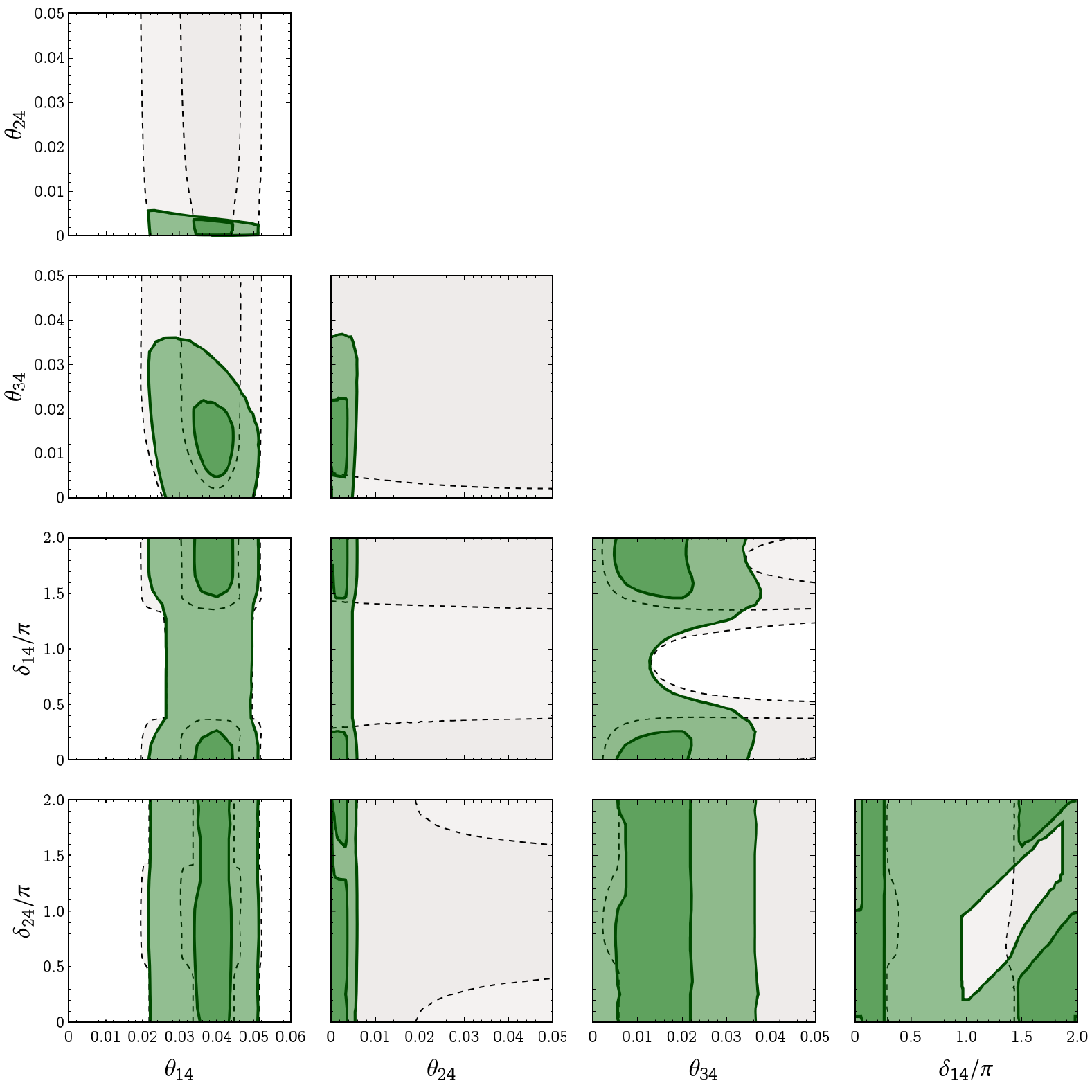}
  \caption{
  Regions of parameter space consistent with the constraints on $|V_{ij}|$ and $\gamma$ (dashed lines) and additionally consistent with the bounds on meson mixing, $|\epsilon_K|$ and the requirement of perturbativity (solid lines). For both sets of constraints, $2\sigma$ ($3\sigma$) contours are shown in darker (lighter) colour. }
  \label{fig:correlations}
\end{minipage}}
\end{figure}
\clearpage
}

\begin{figure}[t!]
  \centering
  \includegraphics[width=0.82\textwidth]{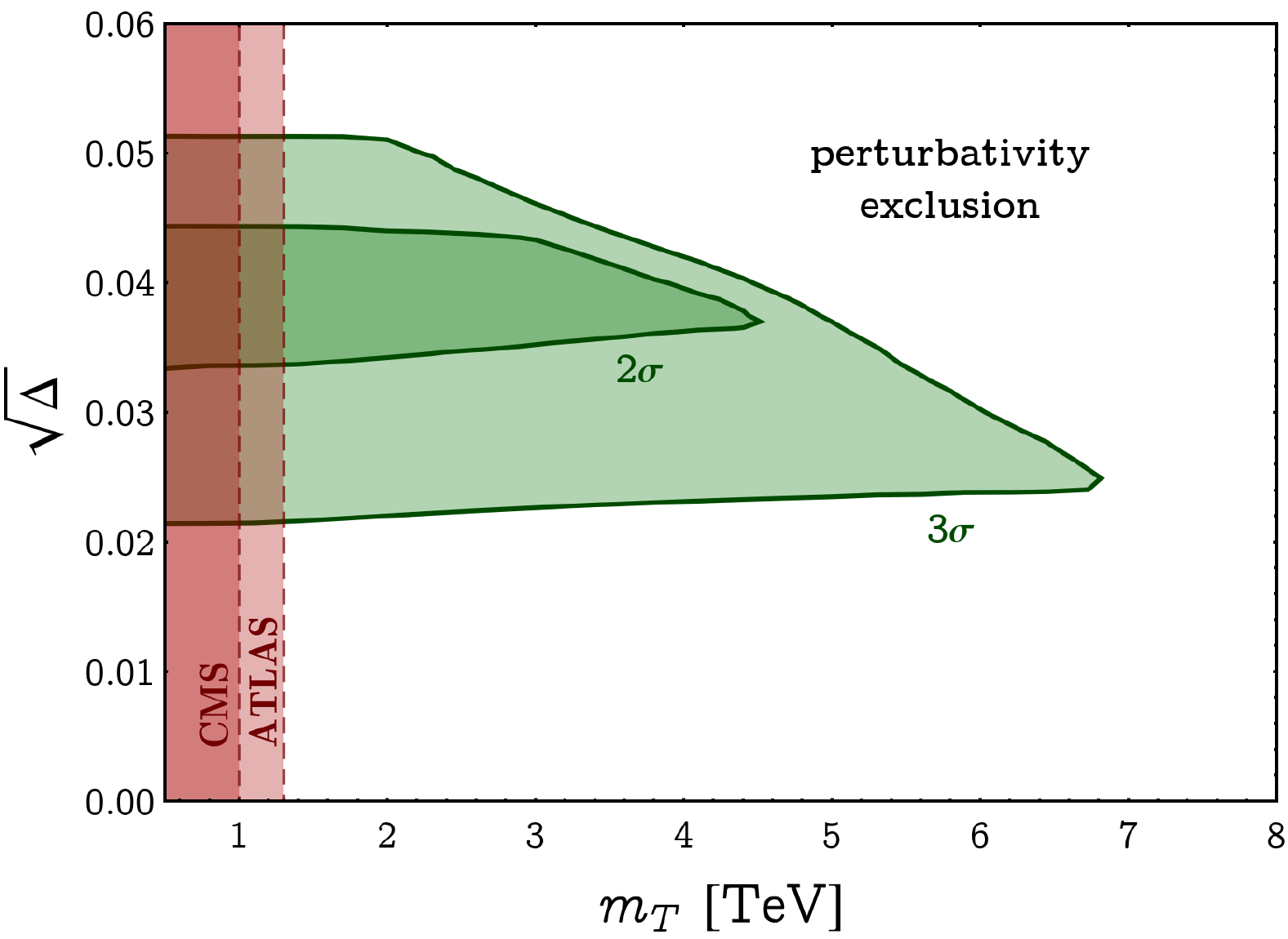}\qquad${}$
  \caption{
   Masses $m_T$ of the new heavy quark and first-row deviations from unitarity ($\sqrt{\Delta} \simeq \theta_{14}$) compatible with the bounds on $|V_{ij}|$, $\gamma$, meson mixing, $|\epsilon_K|$ and with the requirement of perturbativity. The latter constraint imposes an upper limit on $m_T$, while lower bounds are set by ATLAS~\cite{Aaboud:2018pii} and CMS~\cite{Sirunyan:2019sza} (95\% CL).
   }
  \label{fig:perturbativity}
\end{figure}

The perturbativity constraint also restricts the allowed values of $m_T$, which are shown against first-row deviations from unitarity in Figure~\ref{fig:perturbativity}. As anticipated in section~\ref{sec:perturbativity}, the maximum value for $m_T$ depends on the size of the deviations from unitarity. Fixing $\sqrt{\Delta} = 0.04$, one finds $m_T \lesssim 5$ TeV. Taking into account the full $3\sigma$ region of the fit, the bound becomes $m_T \lesssim 7$ TeV.

\begin{figure}[t!]
  \centering
\makebox[0pt]{\begin{minipage}{1.15\textwidth}
\!\!\!\!\!\!\!\!\!\!
    \begin{subfigure}[b]{0.5\textwidth}
        \includegraphics[width=\textwidth]{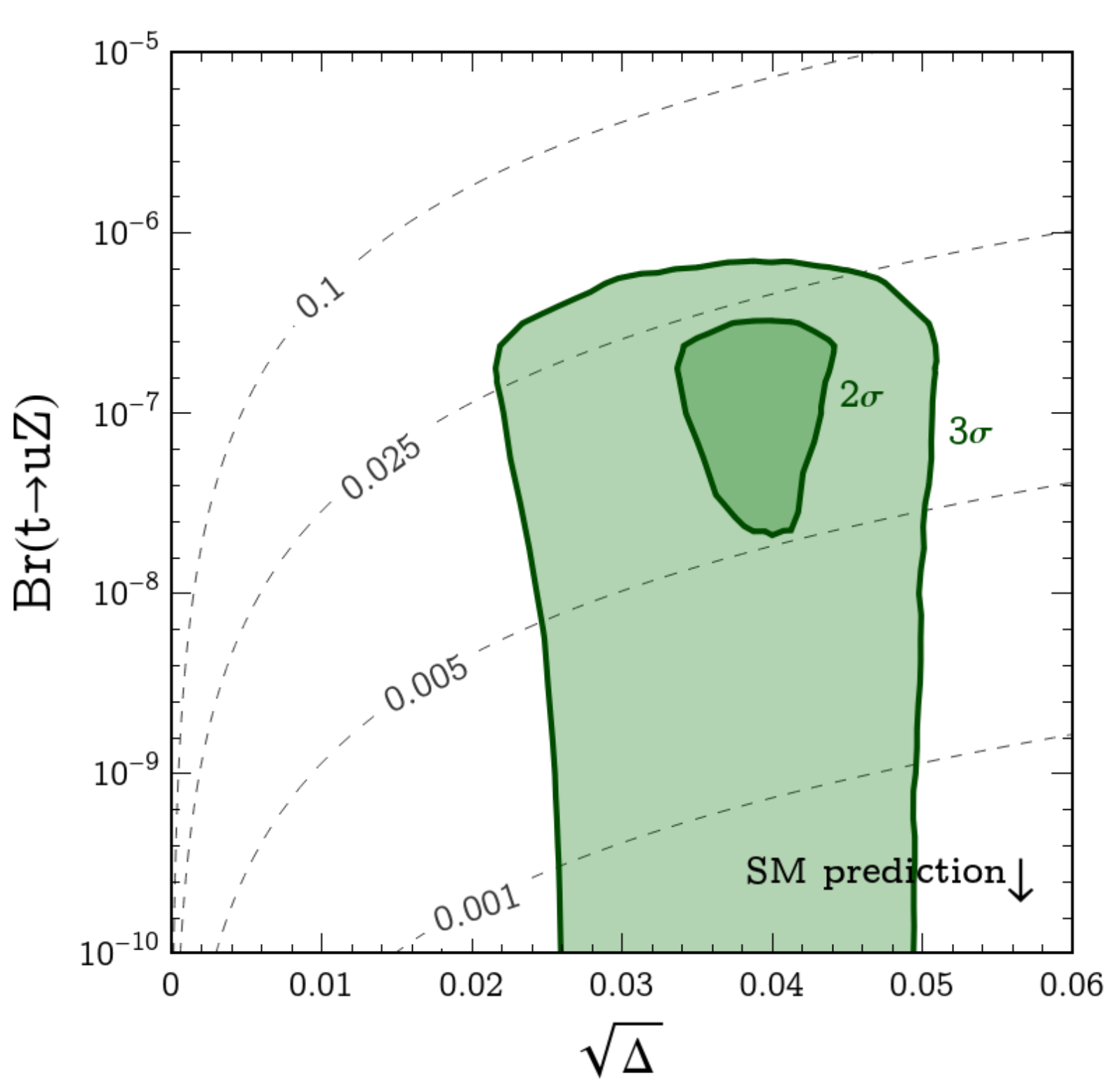}
        \label{fig:tuZ}
    \end{subfigure}\qquad
    \begin{subfigure}[b]{0.5\textwidth}
        \includegraphics[width=\textwidth]{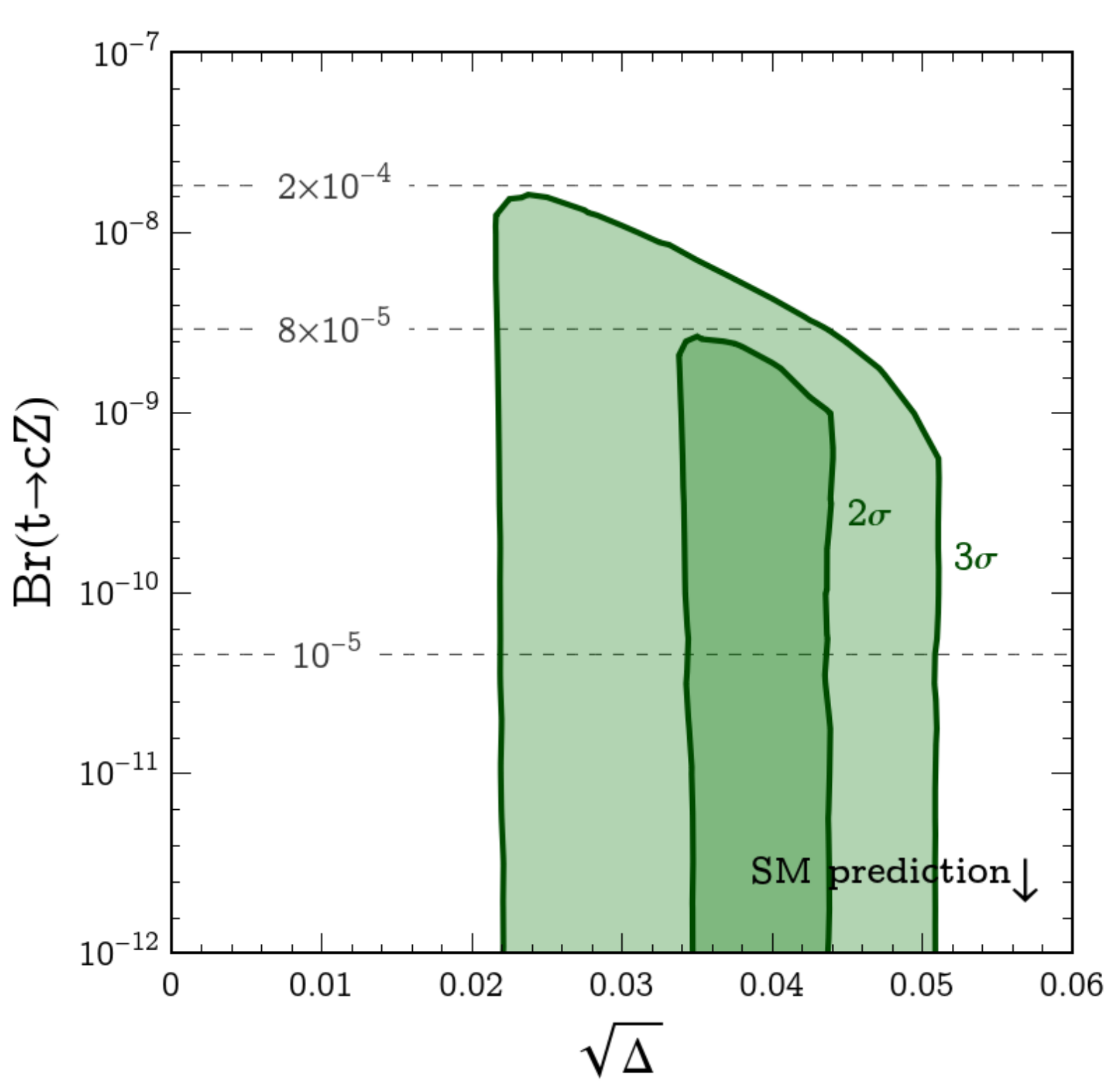}
        \label{fig:tcZ}
    \end{subfigure}
 \caption{Branching ratios for the rare top decays $t \rightarrow u Z$ (left) and $t \rightarrow c Z$ (right) as a function of first-row deviations from unitarity $\Delta$ ($\sqrt{\Delta} \simeq \theta_{14}$). These may considerably exceed the SM expectations Br$(t \to u Z)_\text{SM} \sim 10^{-16}$ and Br$(t \to c Z)_\text{SM} \sim 10^{-14}$~\cite{AguilarSaavedra:2004wm} (not shown). Dashed contours refer to different values for $\theta_{34}$ (left) and for the product $\theta_{24}\,\theta_{34}$ (right), in the small angle approximation.}
  \label{fig:raretop}
\end{minipage}}
\end{figure}

Finally, we present in Figure~\ref{fig:raretop} the predicted values for the branching ratios of rare top decays $t\to uZ$ (left) and $t\to cZ$ (right), as a function of $\sqrt{\Delta}$. These may considerably exceed the SM predictions. One finds $2.0\times 10^{-8} < \text{Br}(t\to uZ) < 3.5 \times 10^{-7}$ at the $2\sigma$ level, which is still 3 orders of magnitude below present bounds (see section~\ref{sec:raretop}). The decay $t\to cZ$ may instead be arbitrarily suppressed, even at the $2 \sigma$ level, by the allowed smallness of $\theta_{24}$.


\vskip 2mm

Note that in the $2\sigma$ fit region we have $|V_{41}| > |V_{42}|, |V_{43}|$, which implies that the up-type vector-like quark $T$ couples more strongly to the up quark than to the charm and top quarks, challenging the typical assumption that the new quark couples only to the third SM quark generation.
In what follows, we describe a low-$\chi^2$ benchmark belonging to this $2\sigma$ region.
The selected point in parameter space corresponds to a mass $m_T = 1.5$ TeV and is described by
\begin{equation}
\begin{split}
    &\theta_{12}= 0.2265 ~,~~\theta_{13} =0.003818 ~,~ \theta_{23}=0.03998 ~,~~\\[1mm]
    &\theta_{14}\simeq \sqrt{\Delta_{(1)}} = 0.03951 ~,~~ \theta_{24}\simeq \sqrt{\Delta_2} = 0.002078 ~,~~ \theta_{34}\simeq \sqrt{\Delta_3}=0.01271 ~,~~\\[1mm]
    &\delta_{13} = 0.396\, \pi ~,~~ \delta_{14} = 1.818\, \pi ~,~~ \delta_{24}= 0.728\, \pi\,,
    \end{split}
\end{equation}
corresponding to $\chi^2 \simeq 3.2$, to a perturbativity factor $p-1 \simeq 0.13$ and to the following values for observables:
\begin{equation}
\begin{split}
    &x_D \simeq 0.18\% ~,~~\Delta m_K \simeq 6.3 \times 10^{-13}~\text{MeV} ~,~~
    \Delta m_B \simeq 3.0 \times 10^{-10}~\text{MeV} ~,~~ \\[1mm]
    &\Delta m_{B_s} \simeq 2.7 \times 10^{-10}~\text{MeV} ~,~~
    |\epsilon_K| \simeq 1.9 \times 10^{-4} ~,~~\\[1mm]
    &\text{Br}(t\to uZ) \simeq 1.2 \times 10^{-7}~,~~ 
    \text{Br}(t\to cZ) \simeq 3.2 \times 10^{-10}~,~~\\[1mm]
    &\alpha \simeq 84.92\degree ~,~~ \beta \simeq 23.85\degree ~,~~ \gamma \simeq 71.23\degree\,,
    \end{split}
\end{equation}
where $\alpha = \arg (-V_{td}V_{ub}V_{tb}^*V_{ud}^*)$ and $\beta = \arg (-V_{cd}V_{tb}V_{cb}^*V_{td}^*)$. The absolute values of the rotation matrix of eq.~\eqref{eq:Vparam} read, for this benchmark,
\begin{equation}
    |\mathcal{V}_L^\dagger| \simeq
  \tikz[baseline=(M.west)]{%
    \node[matrix of math nodes,matrix anchor=west,left delimiter=(,right delimiter=),ampersand replacement=\&] (M) {%
 0.9737 \& 0.2243 \& 0.0038 \& \,0.0395\\
 0.2243 \& 0.9737 \& 0.0400 \& \,0.0021 \\
 0.0081 \& 0.0393 \& 0.9991 \& \,0.0127 \\
 0.0390 \& 0.0065 \& 0.0126 \& \,0.9991 \\
    };
    \node[draw,fit=(M-1-1)(M-3-3),inner sep=-1pt] {};
    \node[draw,dashed,fit=(M-1-1)(M-4-3),inner sep=1pt] {};
  }
\,,
\end{equation}
where $|V|$ is given by the first 3 columns of $|\mathcal{V}_L^\dagger|$ (dashed block) and $|K_\text{CKM}|$ is the $3 \times 3$ upper-left block of $|\mathcal{V}_L^\dagger|$ (unbroken line).
We also give the absolute values of the entries of the matrices $F^u$ and $\eta$ for this benchmark,
\begin{equation}
\begin{split}
    |F^u| &= \begin{pmatrix}
0.99843969 &0.00008203 &0.00050179 &0.03946672\\
 0.00008203 &0.99999569 &0.00002638 &0.00207496\\
 0.00050179 &0.00002638 &0.99983863 &0.01269224\\
 0.03946672 &0.00207496 &0.01269224 &0.001726  
\end{pmatrix}\,,
\\[2mm]
|\eta| &=  \begin{pmatrix}
 0.78016 &0.04102 &0.25089\\ 
 0.04102 &0.00216 &0.01319 \\
 0.25089 &0.01319 &0.08069
 \end{pmatrix}
\times 10^{-3}\,.
\end{split}
\end{equation}
These matrices encode the structure of FCNC and deviations from unitarity respectively and are connected by eq.~\eqref{eq:Wu3}.

Finally, one can write explicitly the mass matrix $\mathcal{M}_u$ for this benchmark in the weak basis where the down-type quark mass matrix is diagonal, $\mathcal{M}_d = \mathcal{D}_d =\diag(m_d,m_s,m_b)$, and in which the upper-left $2 \times 2$ block of $\mathcal{M}_u$ is rotated into the zero matrix,
\begin{equation}
    |\mathcal{M}_u|= \left(
\begin{array}{cccc}
 0 & 0 &  1.39 & 58.57 \\
 0 & 0 & 6.776 & 9.775 \\
 0.001658 & 15.38  & 170.9 & 18.92  \\
0.03206 & 1.486 & 2.166 & 1499
 \\
\end{array}
\right)~\text{GeV}\,.
\label{eq:massmatrix}
\end{equation}
The latter rotation can be achieved solely via transformations from the right. 
In this WB one can read off $(\mathcal{M}_u)_{33} \simeq m_t$, $(\mathcal{M}_u)_{44} \simeq m_T$, and ${(\mathcal{M}_u)_{14}}/{(\mathcal{M}_u)_{44}} \simeq \sqrt{\Delta}$. 
One sees that in this example $(\mathcal{M}_u)_{14}<(\mathcal{M}_u)_{33}$.

\vskip 2mm
Note that if instead of adding an up-type VLQ we chose to add a down-type VLQ to the SM, the down-type quark mass matrix $\mathcal{M}_d$ would share the qualitative features of the $\mathcal{M}_u$ described above.
In particular, in the WB where $\mathcal{M}_u$ is diagonal and the upper-left $2 \times 2$ block of $\mathcal{M}_d$ is transformed into the zero matrix, the structure of $\mathcal{M}_d$ would need to be similar to that of eq.~\eqref{eq:massmatrix} in order to obtain the required deviation from unitarity in the first-row of the CKM matrix.
In such a basis, one expects $(\mathcal{M}_d)_{33} \sim m_b$, with $m_b(M_Z) \simeq 2.89$ GeV~\cite{Xing:2007zj}. On the other hand, experimental bounds on the mass $m_B$ of the new heavy quark in this scenario are close to those on $m_T$, i.e.~$m_B \gtrsim 1$ TeV~\cite{Aaboud:2018pii, Sirunyan:2019sza}, implying a large $(\mathcal{M}_d)_{44} \sim m_B$. The requirement of reproducing the observed deviation from unitarity $\sqrt{\Delta} \simeq 0.04$ would force $(\mathcal{M}_d)_{14} \gtrsim 50~\text{GeV} \gg (\mathcal{M}_d)_{33}$. Since the terms in the first three rows of the mass matrix share a common origin in electroweak symmetry breaking, this hierarchy between $(\mathcal{M}_d)_{14}$ and $(\mathcal{M}_d)_{33}$ may not be appealing or plausible in a theory of flavour addressing the gap $m_b \ll m_t$ in third-generation quark masses.

\section{Summary and Conclusions}
\label{sec:conclusions}

Lately, there has been a lot of interest in models with vector-like quarks (for a review see~\cite{Aguilar-Saavedra:2013qpa} and references therein).
Hints have recently emerged pointing towards the violation of unitarity in the first row of the quark mixing matrix. Such deviations from unitarity arise naturally in scenarios with vector-like quarks. The addition of a down-type VLQ to the SM can bring about a direct effect on the first row of the CKM. In this paper we emphasize that such an effect is also generically present when extending the SM by a single up-type VLQ. 

The addition of vector-like quarks is one of the simplest and most plausible  extensions of the Standard Model. This addition violates the dogma requiring the absence of tree-level $Z$-mediated flavour changing neutral currents (FCNC).
In particular, the introduction of an isosinglet vector-like $Q=2/3$ quark leads to $Z$-mediated up-sector tree-level contributions to processes like $D^0$-$\overline{D}^0$ mixing. These FCNC are naturally suppressed in this framework. The suppression mechanism results from the presence of two mass scales, the scale $v$ of electroweak symmetry breaking and the scale associated to the bare mass terms of the vector-like quarks. The latter mass terms are gauge-invariant and therefore they can be significantly larger than the electroweak scale.

We have shown that it is possible to produce a deviation from unitarity in the first row of the CKM matrix in agreement with the present experimental hint, while at the same time respecting the stringent experimental constraints 
arising from $D^0$-$\overline{D}^0$, $K^0$-$\overline{K}^0$, $B_{d,s}^0$-$\overline{B}_{d,s}^0$ mixings.
We have also seen that one can translate the requirement of having perturbative Yukawa couplings into a bound on the mass $m_T$ of the vector-like quark. The new quark $T$ is required to have a mass lower than about 7 TeV, while if one fixes $\sqrt{\Delta}= 0.04$ then $m_T \lesssim 4.4$ TeV.
In section~\ref{sec:phenomenology}, we have explicitly presented a benchmark with $m_T = 1.5$ TeV. Inspection of the quark mass matrices suggests that the up-type VLQ solution to the hinted unitarity problem may be more appealing than the down-type one, in light of the required hierarchies between mass matrix elements.

If the violations of $3\times 3$ unitarity in the CKM are experimentally confirmed, they may be the first sign of a novel class of New Physics at the reach of LHC and its upgrades.

\vskip 1cm
{\it Note added:} Just before this work was ready to be sent for publication we came across a new paper that appeared recently in the arXiv~\cite{Belfatto:2021jhf}, which has some overlap with our work.  However, significant new material is presented here and our discussion follows a somewhat different approach.

\section*{Acknowledgements}

We would like to thank Bill Marciano for an interesting conversation which gave us 
a new insight into possible unitarity violations, from the point of view of an expert on the subject.
We also thank Francisco Botella for valuable comments.
This work was partially supported by Fundação para a Ciência e a Tecnologia (FCT, Portugal) through the projects
CFTP-FCT Unit 777 (UIDB/00777/2020 and UIDP/00777/2020),
PTDC/FIS-PAR/29436/2017, CERN/FIS-PAR/0004/2019 and 
CERN/FIS-PAR/0008/2019,
which are partially funded through POCTI (FEDER), COMPETE, QREN and EU. P.M.F.P.~acknowledges support from FCT through the PhD grant SFRH/BD/145399/2019.
G.C.B.~and M.N.R.~benefited from discussions that were prompted through the HARMONIA project of the National Science Centre, Poland, under contract UMO-2015/18/M/ST2/00518 (2016-2019), which has been extended.

\appendix

\bibliographystyle{JHEPwithnote}
\bibliography{bibliography}

\end{document}